\documentclass[aps,prb,showpacs]{revtex4}%
\usepackage{amsmath}
\usepackage{amsfonts}
\usepackage{amssymb}
\usepackage{graphicx}
\usepackage{latexsym}
\usepackage{tabularx}
\usepackage{hhline}%
\begin{document}
\title{Ab-initio study of the electric transport in gold nanocontacts}

\author{K.\ Palot\'as$^{1}$, B.\ Lazarovits$^{1}$, L.\ Szunyogh$^{1,2}$,
P.\ Weinberger$^{1}$}

\bigskip
\affiliation{$^{1}$Center for Computational Materials Science, 
Vienna University of Technology, \linebreak 
Gumpendorferstr.\ 1.a., A-1060 Vienna, Austria \linebreak$^{2}%
$Department of Theoretical Physics, Center for Applied Mathematics and
Computational Physics, Budapest University of Technology and Economics,
Budafoki \'ut 8., H-1521 Budapest, Hungary}

\bigskip
\date{Mar 10, 2004, submitted to PRB}

\begin{abstract}
By employing a real-space formulation of the Kubo-Greenwood equation based on
a Green's function embedding technique combined with the fully relativistic
spin-polarized Korringa-Kohn-Rostoker method a detailed investigation of the
electrical transport through atomic-scaled contacts between two Au(001)
semi-infinite systems is presented. Following a careful numerical test of the
method the conductance of Au nanocontacts with different geometries is
calculated. In particular, for a contact formed by a linear chain of Au atoms
a conductance near 1 $G_{0}$ is obtained. The influence of transition metal
impurities (Pd, Fe and Co) placed on various positions near the center of a
particular contact is also studied. We found that the conductance is very
sensitive to the position of the magnetic impurities and that the mechanism
for the occurring relative changes can mainly be attributed to the impurities'
minority $d$-band inducing resonant line-shapes in the $s$-like DOS at the
center of the contact.

\end{abstract}
\maketitle

\parskip 5pt

\section{Introduction}

The number of theoretical and experimental investigations of electronic
structure and transport properties of atomic-sized conductors has greatly been
increased over the last decade. The increasing interest for investigating
atomic-sized conductors is driven by the possibility of using such systems in
future nanoelectronic technologies. Widely applied methods for fabricating
nanocontacts between macroscopic electrodes are the mechanically controllable
break junction (MCBJ) technique \cite{mcbj1,mcbj2,Smit,halbritter} and
scanning tunneling microscopy (STM) \cite{crommie,gimzewski,brandbyge} by
pushing the tip intentionally into the surface. The crucial problems for both
methods are the presence of contaminants and the mechanical stability. At
sufficiently low temperatures the measurements revealed a quantized
conductance for atomic sized nanocontacts made of various materials, not only
pure metals but also alloys \cite{AuPd,heemskerk}. Nanocontacts made of gold
are presumably the most studied systems in the literature both theoretically
and experimentally. A dominant peak very close to the conductance quantum, 1
$G_{0}=2e^{2}/h$, has been reported for gold (and other noble metals) in the
conductance histogram \cite{mcbj2,brandbyge}, attributed to the highly
transmitting $sp$--channel across a linear chain connecting the two
electrodes. It was also found that the chain formation is in close connection
with surface reconstruction phenomena \cite{Smit}. For a comprehensive review
of the field of atomic-sized conductors, see Ref.~\onlinecite{RuitRev}.

In order to understand the mechanism of nanocontact formation, electronic
structure and transport, different theoretical methods have been developed.
Some theoretical studies use tight--binding methods \cite{brandbyge1,solanki},
others are based on ab initio density functional theory
\cite{stepanyuk,mertig,opitz}. Most of the transport studies rely on the
Landauer-B\"uttiker approach \cite{landauer,buttiker}, although Baranger and
Stone adopted the more sophisticated Kubo-Greenwood formula
\cite{kubo,greenwood,luttinger,butler} for calculating the conductance between
free electron leads \cite{baranger}. By using this approach a recent study
\cite{mertig} focused on the effect of transition metal imperfections inserted
into an infinite Cu wire showing that the conductance of the wire decreased
due to the different conductance for the two spin channels (spin-filter effect).

The fully relativistic screened Korringa-Kohn-Rostoker (SKKR) Green's function
method proved to be an effective method to calculate electronic structure and
magnetic properties of layered systems \cite{skkr,rsp-skkr}. Combined with an
embedding technique based on multiple scattering theory, calculations have
been performed for magnetic clusters on surfaces \cite{bence1,bence2,bence3}.
Employing the Kubo-Greenwood formula within this method permits one to
investigate transport properties of atomic sized structures \cite{palotas}. In
this paper we report on calculations of transport properties of gold
nanocontacts. We first briefly review the theoretical background of the
applied method and give numerical evidence of its reliability. We then
calculate and analyze the conductance for different atomic arrangements
between semi-infinite Au(001) systems and investigate the effect of transition
metal impurities on the conductance. We find a qualitatively satisfactory
explanation of the observed changes in the conductance in terms of changes in
the $s$-like local density of states (LDOS) at the center of the point contact
caused by interactions with the $d$-like states of the impurity.

\section{Theoretical approach}

\subsection{Expression of the conductivity}


The static limit of the optical conductivity tensor is given by the
Kubo-Luttinger formula \cite{kubo,luttinger} as
\begin{align}
\label{eq:start}\sigma_{\mu\nu}  &  =-\frac{\hbar}{2\pi V}\int_{-\infty
}^{\infty}d\varepsilon\,f(\varepsilon)\;\\
&  \times Tr\left\{  J_{\mu} \frac{d G^{+}(\varepsilon)}{d\varepsilon} J_{\nu
}[G^{+}(\varepsilon)-G^{-}(\varepsilon)] - J_{\mu}[G^{+}(\varepsilon
)-G^{-}(\varepsilon)] J_{\nu}\frac{d G^{-}(\varepsilon)}{d\varepsilon
}\right\}  \;,\nonumber
\end{align}
where V is the volume of the system, $f(\varepsilon)$ is the Fermi-Dirac
distribution function, $J_{\mu}$ is the $\mu$th component ($\mu=x,y$ or $z$)
of the current density operator, $G^{\pm}(\varepsilon)$ are the corresponding
(upper or lower) side limits of the resolvent of an appropriate
Kohn-Sham(-Dirac) Hamiltonian, while $Tr$ denotes the trace of an operator.
Integration by parts yields
\begin{equation}
\sigma_{\mu\nu}=-\int_{-\infty}^{\infty}d\varepsilon\,\frac{df(\varepsilon
)}{d\varepsilon}S_{\mu\nu}\left(  \varepsilon\right)  \;,
\end{equation}
with%
\begin{align}
S_{\mu\nu}\left(  \varepsilon\right)   &  =-\frac{\hbar}{2\pi V}\int_{-\infty
}^{\varepsilon}d\varepsilon^{\prime}\,\nonumber\\
&  \times Tr\left\{  J_{\mu}\frac{d G^{+}(\varepsilon^{\prime})}
{d\varepsilon^{\prime}} J_{\nu}[G^{+}(\varepsilon^{\prime})-G^{-}
(\varepsilon^{\prime})]- J_{\mu}[G^{+}(\varepsilon^{\prime})-G^{-}%
(\varepsilon^{\prime})] J_{\nu}\frac{d G^{-}(\varepsilon^{\prime}%
)}{d\varepsilon^{\prime}}\right\}  \;,
\end{align}
which has the meaning of a zero-temperature, energy dependent conductivity.
For $T=0,$ $\sigma_{\mu\nu}$ is obviously given by
\begin{equation}
\sigma_{\mu\nu}=S_{\mu\nu}\left(  \varepsilon_{F}\right)  \;.
\end{equation}
A numerically tractable formula can be obtained only for the \emph{diagonal
elements} of the conductivity tensor, namely,%
\begin{align}
&  Tr\left\{  J_{\mu}\frac{d G^{+}(\varepsilon)}{d\varepsilon}J_{\mu}%
[G^{+}(\varepsilon)-G^{-}(\varepsilon)]-J_{\mu}[G^{+}(\varepsilon
)-G^{-}(\varepsilon)]J_{\mu}\frac{d G^{-}(\varepsilon)}{d\varepsilon}\right\}
=\nonumber\\
&  =Tr\left\{  J_{\mu}\frac{d}{d\varepsilon}\left[  G^{+}(\varepsilon
)-G^{-}(\varepsilon)\right]  J_{\mu}[G^{+}(\varepsilon)-G^{-}(\varepsilon
)]\right\} \nonumber\\
&  =\frac{1}{2}\frac{d}{d\varepsilon}Tr\left\{  J_{\mu}[G^{+}(\varepsilon
)-G^{-}(\varepsilon)]J_{\mu}[G^{+}(\varepsilon)-G^{-}(\varepsilon)]\right\}
\;,
\end{align}
yielding the widely used Kubo-Greenwood formula \cite{greenwood,butler} of the
dc-conductivity at finite temperatures,
\begin{equation}
\label{eq:Kubo}\sigma_{\mu\mu}=-\frac{\hbar}{4\pi V} \int_{-\infty}^{\infty
}d\varepsilon\, \left(  - \frac{df(\varepsilon)}{d\varepsilon} \right)
Tr\left\{  J_{\mu}[G^{+}(\varepsilon)-G^{-}(\varepsilon)] J_{\mu}%
[G^{+}(\varepsilon)-G^{-}(\varepsilon)]\right\}  \; .
\end{equation}
On the other hand, Eq.\ (\ref{eq:start}) can be reformulated as follows,
\begin{align}
\label{eq:Baranger}\sigma_{\mu\nu}  &  =\frac{\hbar}{2\pi V}\int_{-\infty
}^{\infty}d\varepsilon\,f(\varepsilon)\;Tr\left\{  J_{\mu}\frac{d
G^{+}(\varepsilon)}{d\varepsilon} J_{\nu}G^{-}(\varepsilon)+ J_{\mu}%
G^{+}(\varepsilon) J_{\nu} \frac{d G^{-}(\varepsilon)}{d\varepsilon}\right\}
\nonumber\\
&  -\frac{\hbar}{2\pi V}\int_{-\infty}^{\infty}d\varepsilon\,f(\varepsilon
)\;Tr\left\{  J_{\mu}\frac{d G^{+}(\varepsilon)}{d\varepsilon} J_{\nu}%
G^{+}(\varepsilon)+ J_{\mu}G^{-}(\varepsilon) J^{\nu}\frac{d G^{-}%
(\varepsilon)}{d\varepsilon}\right\} \nonumber\\
& \nonumber\\
&  =\frac{\hbar}{2\pi V}\int_{-\infty}^{\infty}d\varepsilon\,\left(
-\frac{df(\varepsilon)}{d\varepsilon}\right)  Tr\left\{  J_{\mu}%
G^{+}(\varepsilon) J_{\nu}G^{-}(\varepsilon)\right\} \\
&  -\frac{\hbar}{2\pi V}\int_{-\infty}^{\infty}d\varepsilon\,f(\varepsilon
)\;Tr\left\{  J_{\mu}\frac{d G^{+}(\varepsilon)}{d\varepsilon} J_{\nu}%
G^{+}(\varepsilon)+ J_{\mu}G^{-}(\varepsilon) J_{\nu}\frac{d G^{-}%
(\varepsilon)}{d\varepsilon}\right\}  \; ,\nonumber
\end{align}
namely in terms of an equation which is similar to the formulation of Baranger
and Stone \cite{baranger} but clearly can be cast into a relativistic form.

\subsection{Expression of the conductance}

Linear response theory applies to an arbitrary choice for the perturbating
electric field because the response function is obtained in the zero limit of
perturbation. Let us assume, therefore, that a constant electric field,
$E_{z}^{J}$, pointing along the $z$ axis, i.e., normal to the planes, is
applied in all cells of layer $J$. Denoting the $z$ component of current
density averaged over cell $i$ in layer $I$ by $j_{z}^{iI}$, the microscopic
Ohm's law reads as
\begin{equation}
j_{z}^{iI}=\frac{1}{V_{at}}\sum_{j}\sigma_{zz}^{iI,jJ}E_{z}^{J}%
\;,\label{eq:jz}%
\end{equation}
where $V_{at}$ is the volume of the unit cell in layer $I$. Note, that in
neglecting lattice relaxations, $V_{at}$ is uniform in the whole system.
According to the Kubo-Greenwood formula, Eq.\ (\ref{eq:Kubo}), the $zz$
component of the non-local conductivity tensor, $\sigma_{zz}^{iI,jJ}$ can be
written at zero temperature as
\begin{align}
\sigma_{zz}^{iI,jJ} &  =-\frac{\hbar}{4\pi}\int_{\Omega_{iI}}d^{3}r_{iI}%
\int_{\Omega_{jJ}}d^{3}r_{jJ}^{\prime}\\
&  \times Tr\left(  J_{z}[G^{+}(\varepsilon_{F};\mathbf{r}_{iI},\mathbf{r}%
_{jJ}^{\prime})-G^{-}(\varepsilon_{F};\mathbf{r}_{iI},\mathbf{r}_{jJ}^{\prime
})]J_{z}[G^{+}(\varepsilon_{F};\mathbf{r}_{jJ}^{\prime},\mathbf{r}_{iI}%
)-G^{-}(\varepsilon_{F};\mathbf{r}_{jJ}^{\prime},\mathbf{r}_{iI})]\right)
.\nonumber
\end{align}
Here the integration is carried out over the $i$th unit cell in layer $I$,
$\Omega_{iI}$, and the $j$th unit cell in layer $J$, $\Omega_{jJ}$, while $Tr$
denotes a trace over four-component spinors. The total current flowing through
layer $I$ can be written as
\begin{equation}
I_{tot}=A_{\parallel}\sum_{i}j_{z}^{iI}=gU\;,\label{eq:I}%
\end{equation}
where the applied voltage $U$ is
\begin{equation}
U=E_{z}^{J}d_{\perp}\;,\label{eq:U}%
\end{equation}
and $A_{\parallel}$ and $d_{\perp}$ denote the area of the 2D unit cell and
the interlayer spacing, respectively ($V_{at}=A_{\parallel}\,d_{\perp}$).
Combining Eqs.\ (\ref{eq:jz}),(\ref{eq:I}) and (\ref{eq:U}) results in an
expression for the conductance,
\begin{equation}
g=\frac{1}{d_{\perp}^{2}}\sum_{i}\sum_{j}\sigma_{zz}^{iI,jJ},
\end{equation}
where the summations should, in principle, be carried out over all the cells
in layers $I$ and $J$. An alternative choice of the non-local conductivity
tensor is given by Eq.\ (\ref{eq:Baranger}). This leads to a huge
simplification for the conductance because, as shown by Baranger and Stone
\cite{baranger} for free electron leads, the second term appearing in
Eq.\ (\ref{eq:Baranger}) becomes identically zero when integrated over the
layers, $I\neq J$. It should be noted that recently Mavropoulos \emph{et al.}
\cite{mavropoulos} rederived this result by assuming Bloch boundary conditions
for the leads. The conductance can thus be written as
\begin{equation}
g=\frac{\hbar}{2\pi d_{\perp}^{2}}\sum_{i}\sum_{j}\int_{\Omega_{iI}}%
d^{3}r_{iI}\int_{\Omega_{jJ}}d^{3}r_{jJ}^{\prime}Tr\left[  J_{z}%
G^{+}(\varepsilon_{F};\mathbf{r}_{iI},\mathbf{r}_{jJ}^{\prime})J_{z}%
G^{-}(\varepsilon_{F};\mathbf{r}_{jJ}^{\prime},\mathbf{r}_{iI})\right]
\;.\label{eq:cond1}%
\end{equation}
It has to be emphasized that because of the use of linear response theory and
current conservation, the choice of layers $I$ and $J$ is arbitrary in the
above formula. The numerical test of the method will clearly demonstrate this
feature (see Section III.). On the other hand, as shown in
Ref.~\onlinecite{mavropoulos}, when the layers $I$ and $J$ are asymptotically
far from each other, the present formalism naturally recovers the
Landauer-B\"{u}ttiker approach \cite{landauer,buttiker}.

\subsection{Computational details}

Using the embedding technique of multiple scattering the matrix representation
of the scattering path operator (SPO) of a given cluster, $\underline
{\underline{\tau}}_{clus}(\varepsilon)= \left\{  \underline{\tau}%
^{ij}(\varepsilon) \right\}  = \left\{  \tau^{ij}_{QQ^{\prime}}(\varepsilon)
\right\} $, with $i$ and $j$ denoting sites in the cluster, $Q$ and
$Q^{\prime}$ indexing angular momentum quantum numbers, can be expressed as
\cite{bence1}
\begin{equation}
\label{eq:embedding}\underline{\underline{\tau}}_{clus}(\varepsilon
)=\underline{\underline{\tau}}_{host}(\varepsilon)\left[ \underline
{\underline{I}}-\left( \underline{\underline{t}}_{host}^{-1}(\varepsilon
)-\underline{\underline{t}}_{clus}^{-1}(\varepsilon)\right) \underline
{\underline{\tau}}_{host}(\varepsilon)\right] ^{-1} \; ,
\end{equation}
where $\underline{\underline{t}}_{host}(\varepsilon)$, $\underline
{\underline{t}}_{clus}(\varepsilon)$ stand for the corresponding single-site
t-matrices of the host medium and the cluster, and $\underline{\underline
{\tau}}_{host}(\varepsilon)$ for the host-SPO. For a two-dimensional (2D)
translational invariant host medium, the latter one is calculated by
\begin{equation}
\label{eq:BZ-integration}\underline{\tau}_{host}^{iI,jJ}(\varepsilon)=
\frac{1}{\Omega_{BZ}}\int_{BZ} d^{2}k_{\parallel} e^{i\mathbf{k}_{\parallel
}(\mathbf{T}_{j}-\mathbf{T}_{i})} \underline{\tau}_{host}^{IJ}(\varepsilon
,\mathbf{k}_{\parallel}) \; ,
\end{equation}
where $\mathbf{T}_{i}$ and $\mathbf{T}_{j}$ are 2D lattice vectors and the
integral is performed over the 2D Brillouin zone of area $\Omega_{BZ}$.

The self-consistent calculations for both the host and the finite clusters
were performed within the local spin-density approximation (LSDA)
\cite{Vosko}, by using the atomic sphere approximation (ASA) and $l_{max}=2$
for the angular momentum expansion. The semi-infinite host system was
evaluated in terms of the screened Korringa-Kohn-Rostoker method (SKKR)
\cite{skkr,rsp-skkr} by sampling 45 $k_{\parallel}$ points in the irreducible
(1/8) part of the fcc(001) Brillouin zone, see Eq. (\ref{eq:BZ-integration}),
and 16 energy points along a semi-circular contour in the upper complex energy
semi-plane. The latter set-up also applied for the self-consistent cluster
calculations, whereby a sufficiently large number of atoms from the
neighboring host (including sites in the vacuum) was taken into account in
order to serve as a buffer for charge fluctuations. In the case of magnetic
impurities, the orientation of magnetization was chosen to be normal to the
fcc(001) planes (direction $z$). Additional calculations of the magnetic
anisotropy energy confirmed this choice.

In terms of the SPO the conductance in Eq.\ (\ref{eq:cond1}) can be calculated
as
\begin{equation}
g=\frac{\hbar}{2\pi d_{\perp}^{2}}\sum_{i}\sum_{j}tr\left[  \underline{J}%
_{z}^{iI}(\varepsilon_{F}^{-},\varepsilon_{F}^{+})\,\underline{\tau}%
_{clus}^{iI,jJ}(\varepsilon_{F}^{+})\,\underline{J}_{z}^{jJ}(\varepsilon
_{F}^{+},\varepsilon_{F}^{-})\,\underline{\tau}_{clus}^{jJ,iI}(\varepsilon
_{F}^{-})\right]  ,\label{eq:cond}%
\end{equation}
where $\underline{J}_{z}^{iI}(\varepsilon,\varepsilon^{\prime})$ stands for
the relativistic angular momentum representation of the current density
operator in cell $i$ of layer $I$ (see, e.g., in Ref. \onlinecite{palotas}),
and, correspondingly, the trace is performed in angular momentum space.
Inherent to the SKKR method, a finite imaginary part, $\delta$, of the Fermi
energy has to be applied, $\varepsilon_{F}^{\pm}=\varepsilon_{F}\pm i\delta$,
which, however, has to be continued to zero in order to ensure current
conservation. Concomitantly, the number of $k_{\parallel}$ points taken in Eq.
(\ref{eq:BZ-integration}) has to be considerably increased. All results
presented in the next Section for the conductance refer to $\delta=1\mu$Ryd.

\begin{figure}[h]
\begin{center}
\includegraphics[width=10cm,clip]{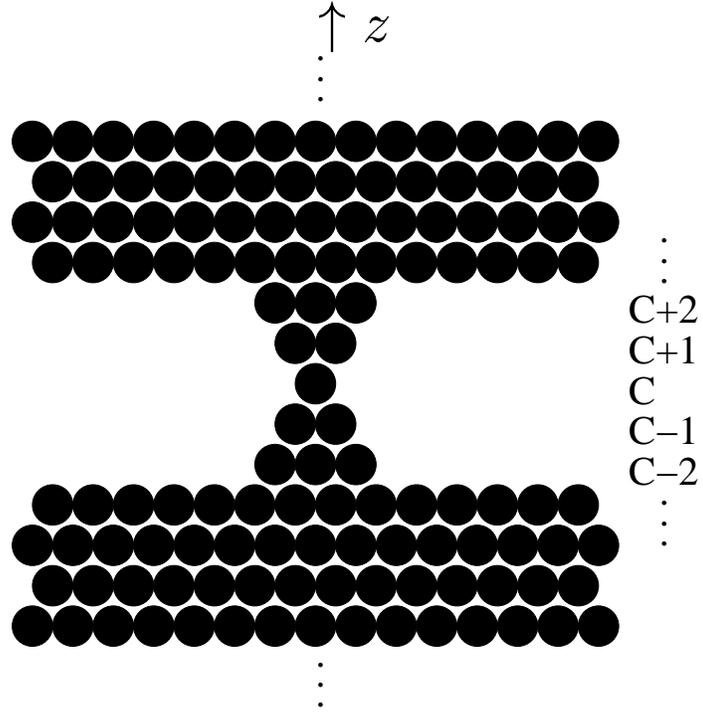}\\[0pt]
\end{center}
\caption{Schematic side view of a point contact between two semi-infinite
leads. The layers are labelled by $C$, {$C \pm1$}, etc.}%
\label{fig:geom-sketch}%
\end{figure}

In the present work no geometry optimization has been carried out, that means
all of the considered sites (both Au, vacuum and impurity sites) correspond to
the positions of an ideal fcc(001) structure of gold with a lattice constant
of $a_{3D}=7.68$ $a.u.$ A schematic view of a typical contact is displayed in
Fig. \ref{fig:geom-sketch}. As follows from the above, atomic sites refer to
layers for which we use the notation: $C$ the \textit{central layer}, $C-1$
and $C+1$ the layers below and above, etc. For the contact shown in Fig.
\ref{fig:geom}a, e.g., the central layer contains 1 Au atom (the rest is built
up from empty spheres), layers $C-1$ and $C+1$ contain 4 Au atoms, layers
$C-2$ and $C+2$ contain 9 Au atoms and, though not shown, all layers $C-n$ and
$C+n$ ($n\geq3$) are completely filled with Au atoms and will be denoted by
full layers.

\section{Results and discussion}

\subsection{Numerical tests on different gold contacts}

\begin{figure}[h]
\begin{center}
\includegraphics[width=15cm,clip]{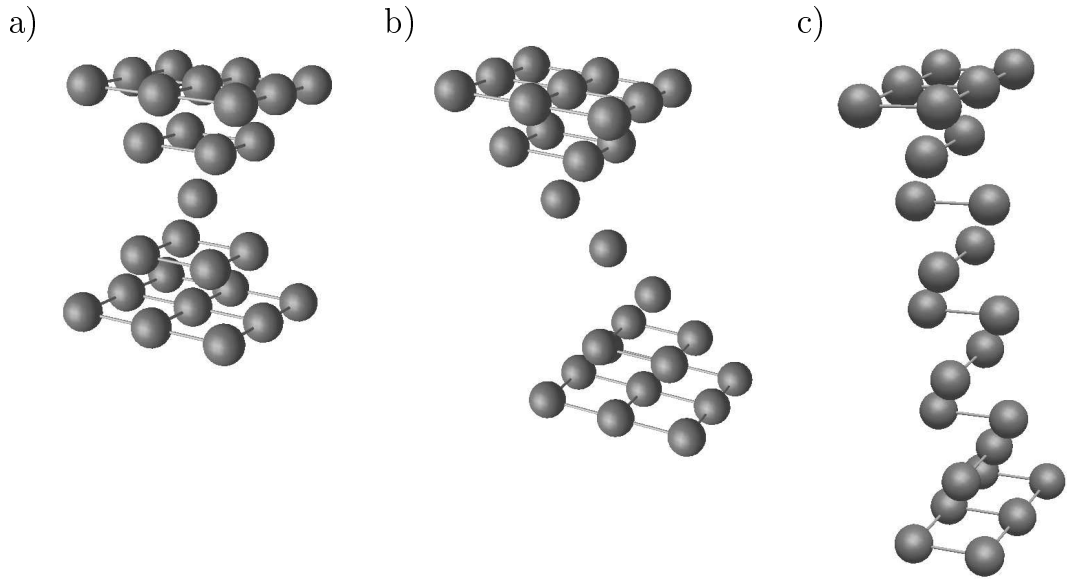}\\[0pt]
\end{center}
\caption{Perspective view of some contacts between two fcc(001) semi-infinite
leads. Only the partially filled layers are shown. a) point contact b) slanted
linear finite chain c) 2x2 finite chain.}%
\label{fig:geom}%
\end{figure}
\begin{figure}[h]
\begin{center}
\includegraphics[width=12cm,clip]{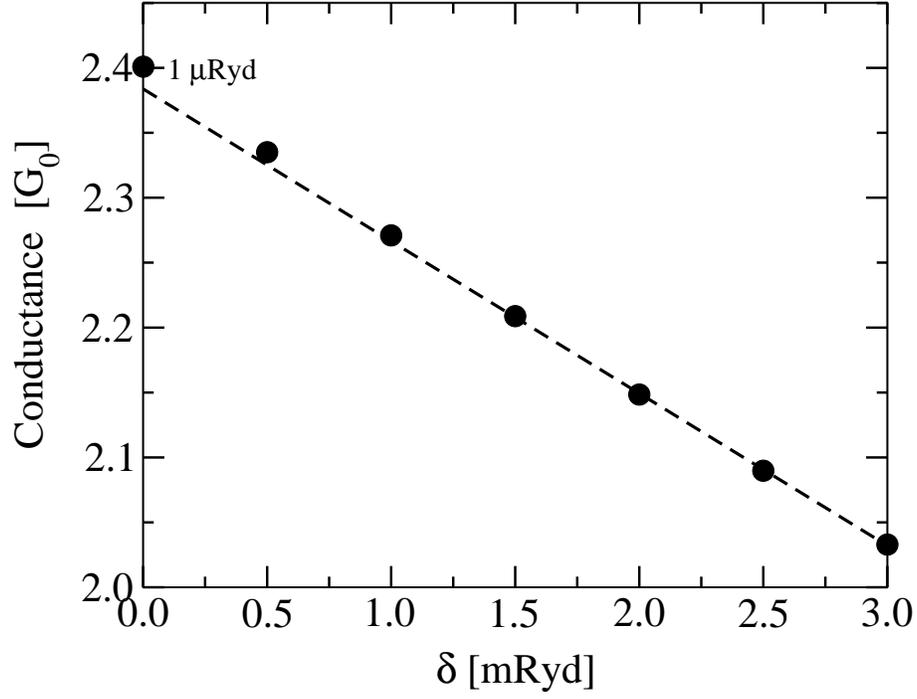}\\[0pt]
\end{center}
\caption{Calculated conductance as a function of the Fermi level broadening,
$\delta$, for the Au contact shown in Fig. \ref{fig:geom}a. The dashed
straight line is a linear fit to the values for $\delta$ = 1.5, 2.0, 2.5, and
3.0 mRyd.}%
\label{fig:delta}%
\end{figure}

As mentioned in Section II a finite Fermi level broadening, $\delta$, has to
be used for the conductance calculations. As an example, for the point contact
depicted in Fig. \ref{fig:geom}a, we investigated the dependence of the
conductance on $\delta$. The summation in Eq.\ (\ref{eq:cond}) was carried out
up to convergence for the first two (symmetric) full layers ($I=C-3,J=C+3$).
As can be seen from Fig. \ref{fig:delta}, the calculated conductances depend
strongly but nearly linear on $\delta$. A straight line fitted for $\delta
\ge1.5$ mRyd intersects the vertical axis at 2.38 $G_{0}$. Assuringly enough,
a calculation with $\delta=1\mu$Ryd resulted in $g=2.40 G_{0}$. Although the
nearly linear dependence of the conductance with respect to $\delta$ enables
an easy extrapolation to $\delta=0$, as what follows all the calculations
refer to $\delta=1 \mu$Ryd.

\begin{figure}[h]
\begin{center}
\includegraphics[width=12cm,clip]{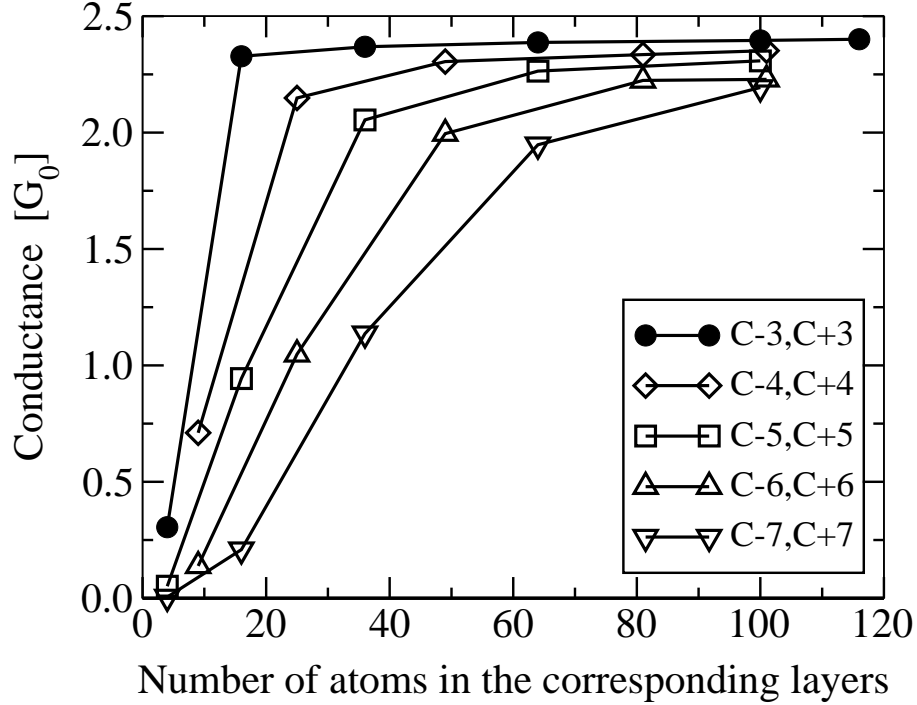}\\[0pt]
\end{center}
\caption{Conductance versus the number of sites included in the sum in Eq.
({\ref{eq:cond}}) for the contact in Fig. \ref{fig:geom}a. The different
curves show conductances as calculated between different pairs of layers. For
a definition of the layer numbering see Fig. \ref{fig:geom-sketch}. }%
\label{fig:conv}%
\end{figure}

For the same type of contact we investigated the convergence of the summation
in Eq.\ (\ref{eq:cond}) over the layers $I$ and $J$, whereby we chose
different symmetric pairs of full layers. The convergence with respect to the
number of atoms in the layers is shown in Fig. \ref{fig:conv}. Convergence for
about 20 atoms can be obtained for the first two full layers ($I=C-3,J=C+3$),
whereas the number of sites needed to get convergent sums gradually increases
if one takes layers farther away from the center of the contact. This kind of
convergence property is qualitatively understandable since the current flows
from the contact within a cone of some opening angle that cuts out sheets of
increasing area from the corresponding layers. As all the calculations were
performed with $\delta=1\mu$Ryd, current conservation has to be expected.
Consequently the calculated conductance ought to be independent with respect
to the layers chosen for the summation in Eq.\ (\ref{eq:cond}). As can be seen
from Fig. \ref{fig:conv} this is satisfied within a relative error of less
then 10\%. It should be noted, however, that for the pairs of layers,
$I=C-n,J=C+n$, $n \ge6$ convergence was not achieved within this accuracy: by
taking more sites in the summations even a better coincidence of the
calculated conductance values for different pairs of layers can be expected.
Fig. \ref{fig:conv} also implies that an application of the
Landauer-B\"uttiker approach to calculate the conductance of nanocontacts is
numerically more tedious than the present one, since, in principle, two layers
situated infinitely far from each other have to be taken in order to represent
the leads.

\begin{figure}[h]
\begin{center}
\includegraphics[width=12cm,clip]{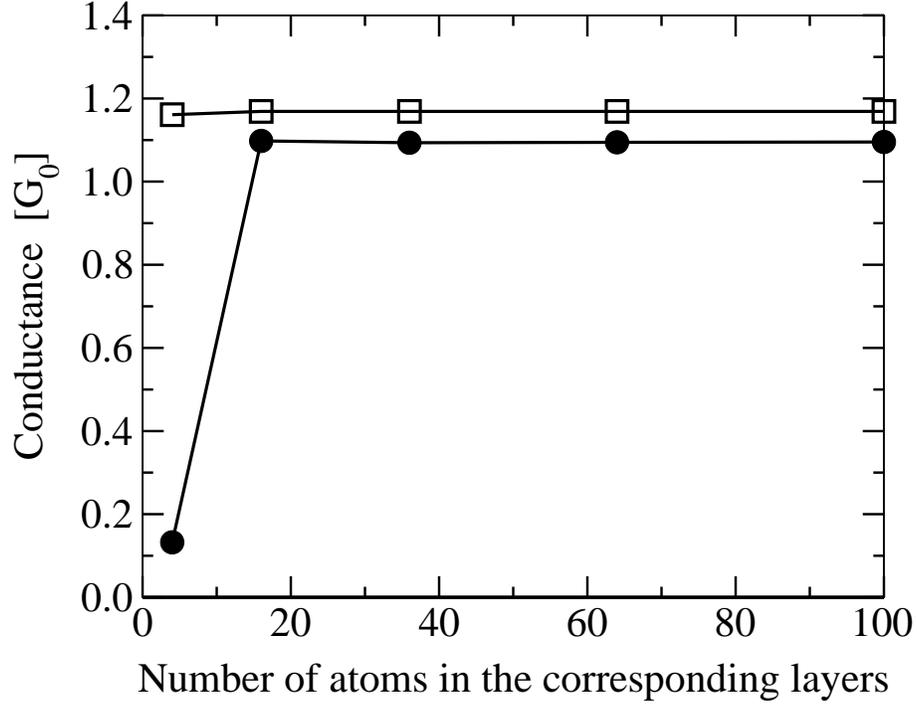}\\[0pt]
\end{center}
\caption{Conductance versus the number of sites included in the sum in Eq.
({\ref{eq:cond}}) for the slanted wire shown in Fig. \ref{fig:geom}b. Full
circles are the results of summing in layers $I=C-4$ and $J=C+4$ (first full
layers), while squares refer to a summation in layers $I=C-2$ and $J=C+2$
(layers containing 4 Au atoms). }%
\label{fig:lincond}%
\end{figure}

Although only one Au atom is placed in the center of the point contact
considered above, see Fig. \ref{fig:geom}a, the calculated conductance is more
than twice as large as the conductance unit. This is easy to understand since
the planes $C-1$ and $C+1$, each containing four Au atoms, are relatively
close to each other and, therefore, tunneling contributes quite a lot to the
conductance through the contact. In order to obtain a conductance around 1
$G_{0}$, detected in the experiments, a linear chain has to be considered. The
existence of such linear chains is obvious from the long plateau of the
corresponding conductance trace with respect to the piezo voltage in the
break-junction experiments. Since, as mentioned in Section II, our method at
present can only handle geometrical structures confined to three dimensional
translational invariant simple bulk parent lattices, as the simplest model of
such a contact we considered a slanted linear chain as shown in Fig.
\ref{fig:geom}b. In here, the middle layer ($C$) and the adjacent layers
($C\pm1$) contain only one Au atom, layers $C\pm2$ and $C\pm3$ 4 and 9 Au
atoms, respectively, while layers $C\pm4$ refer to the first two full layers.
The sum in Eq.\ (\ref{eq:cond}) was carried out for two pairs of layers,
namely for $I=C-4$, $J=C+4$ (full layers) and for $I=C-2 $, $J=C+2$ (not full
layers). The convergence with respect to the number of atoms in the chosen
layers can be seen from Fig. \ref{fig:lincond}. The respective converged
values are 1.10 $G_{0}$ and 1.17 $G_{0}$. In the case of $I=C-2$, $J=C+2$ we
observed that the contribution coming from the vacuum sites is nearly zero:
considering only 4 Au atoms in the summation already gave a value for the
conductance close to the converged one. The small difference between the two
calculated values, 0.07 $G_{0}$, can most likely be attributed to an error
caused by the ASA. Nevertheless, as expected, the calculated conductance is
very near to the ideal value of 1 $G_{0}$.

Another interesting structure is the 2x2 chain described in Ref.
\onlinecite{mertig}. Here we considered a finite length of this structure
sandwiched between two semi-infinite systems, see Fig. \ref{fig:geom}c. The
conductance was calculated by performing the summation for 100 atoms from each
of the first two full layers. As a result we obtained a conductance of 2.58
$G_{0}$. Papanikolaou \emph{et al.} \cite{mertig} got a conductance of 3
$G_{0}$ for an infinite Cu wire to be associated with three conducting
channels within the Landauer approach. For an infinite wire the transmission
probability is unity for all states, therefore, the conductance is just the
number of bands crossing the Fermi level. For the present case of a finite
chain, the transmission probability is less than unity for all the conducting
states. This qualitatively explains the reduced conductance with respect to an
infinite wire.

\begin{figure}[h]
\begin{center}
\includegraphics[width=12cm,clip]{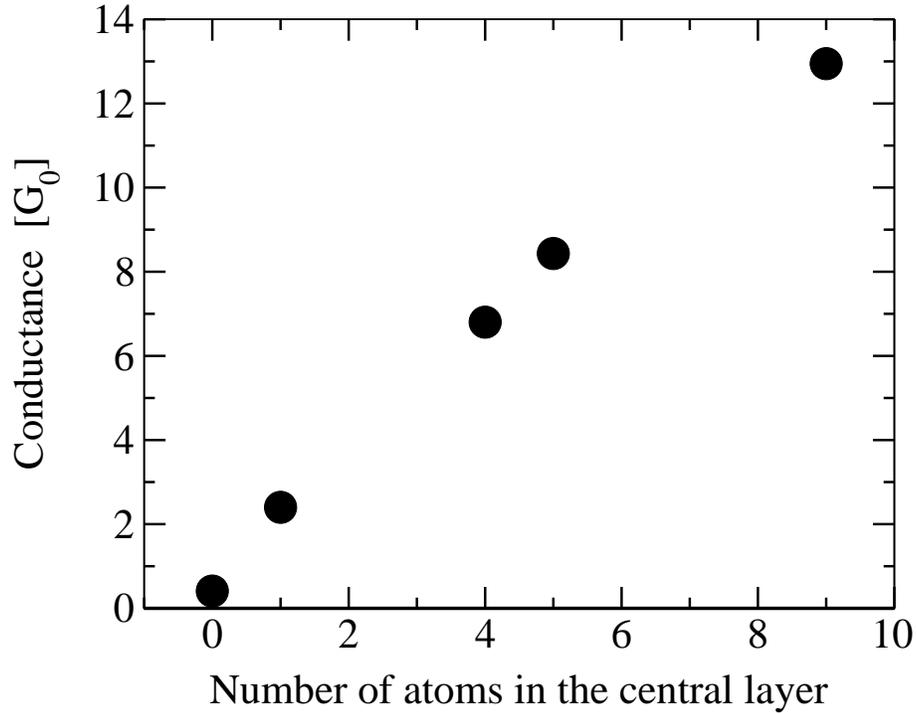}\\[0pt]
\end{center}
\caption{Conductance versus the number of Au atoms in the central layer for
the Au contacts described in Table \ref{tab:geom}.}%
\label{fig:thick}%
\end{figure}

Finally, we studied the dependence of the conductance on the thickness of the
nanocontacts. All the investigated structures have $C_{4v}$ symmetry and the
central layer of the systems is a plane of reflection symmetry. The set-up of
the structures is summarized in Table \ref{tab:geom}. Contact 0 refers to a
broken contact, while the others have different thicknesses from 1 up to 9 Au
atoms in the central layer. In Fig. \ref{fig:thick} the calculated
conductances are displayed as performed by taking nearly 100 atoms from each
of the first two full layers: $I=C-4,J=C+4$ for the broken contact and
$I=C-3,J=C+3$ for all the other cases, see Table \ref{tab:geom}. It can be
seen that the conductance is nearly proportional to the number of Au atoms in
the central layer. This finding can qualitatively be compared with the result
of model calculations for the conductance of a three-dimensional electron gas
through a connective neck as a function of its area in the limit of
$\vartheta_{0}=90^{\circ}$ for the opening angle \cite{Torres}. In the case of
the broken contact, the non-zero conductance can again be attributed to
tunneling of electrons.

\subsection{Gold contact with an impurity}

\begin{figure}[h]
\begin{center}
\includegraphics[width=15cm,clip]{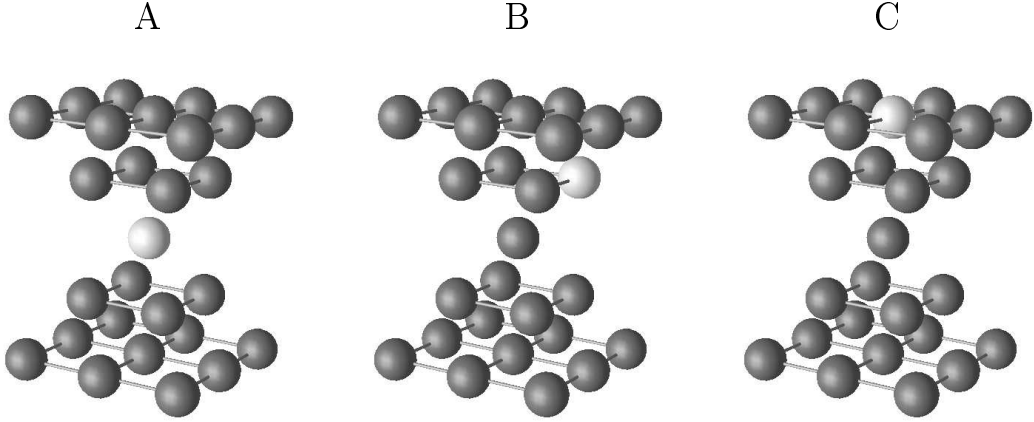}\\[0pt]
\end{center}
\caption{Impurity positions (light grey spheres) in a Au point contact, see
Fig. \ref{fig:geom}a.}%
\label{fig:imp}%
\end{figure}

In recent break junction experiments \cite{AuPd} remarkable changes of the
conductance histograms of nanocontacts formed by AuPd alloys have been
observed when varying the Pd concentration. Studying the effect of impurities
placed into the nanocontact are, in that context, at least relevant for dilute
alloys. The interesting question is whether the presence of impurities can be
observed in the measured conductance. For that reason we investigated
transition metal impurities such as Pd, Fe, and Co placed at various positions
of the point contact as shown in Fig. \ref{fig:geom}a. For the notation of the
impurity positions see Fig. \ref{fig:imp}. The calculated spin and orbital
moments of the magnetic impurities are listed in Table \ref{tab:impmom}. As
usual for magnetic impurities with reduced coordination number\cite{bence1},
both for Fe and Co we obtained remarkably high spin moments, and in all
positions of a Co impurity large orbital moments. In particular, the magnitude
of the orbital moments is very sensitive to the position of the impurity. This
is most obvious in case of Fe, where at positions B and C the orbital moment
is relatively small, but at position A a surprisingly high value of 0.47
$\mu_{B}$ was obtained.

\begin{figure}[h]
\begin{center}
\includegraphics[width=12cm,clip]{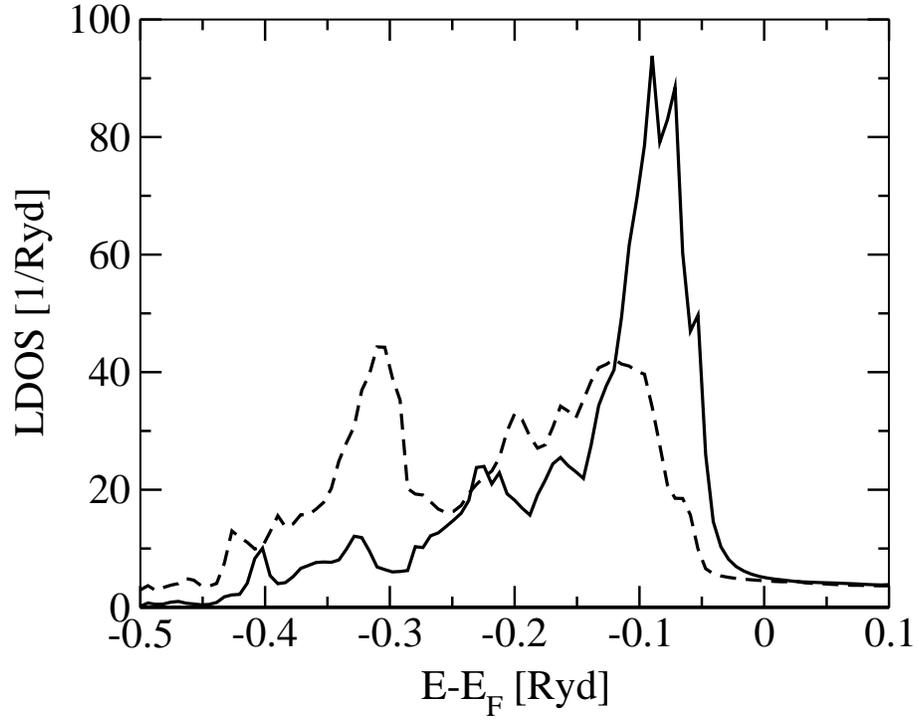}\\[0pt]
\end{center}
\caption{Local density of states of a Pd impurity in position A (solid line)
and in position C (dashed line) of a Au point contact, see Fig. \ref{fig:imp}%
.}%
\label{fig:Pd}%
\end{figure}

The summation over 116 atoms from each of the first two full layers
($I=C-3,J=C+3$) in Eq.\ (\ref{eq:cond}) has been carried out in order to
evaluate the conductance. The calculated values are summarized in Table
\ref{tab:impcond}. A Pd impurity (independent of position) reduces only little
the conductance as compared to a pure Au point contact. This qualitatively can
be understood from the local density of states (LDOS) of the Pd impurity
(calculated by using an imaginary part of the energy of $\delta$ = 1 mRyd). In
Fig. \ref{fig:Pd} we plotted the corresponding LDOS at positions A and C.
Clearly, the change of the coordination number (8 at position A and 12 at
position C), i.e., different hybridization between the Pd and Au $d$ bands,
results into different widths for the Pd $d$-like LDOS. In both cases,
however, the Pd $d$ states are completely filled and no remarkable change in
the LDOS at Fermi level (conducting states) happens.

The case of the magnetic impurities seems to be more interesting. As can be
inferred from Table \ref{tab:impcond}, impurities at position B change only a
little the conductance. Being placed at position A, however, Fe and Co atoms
increase the conductance by 11 \% and 24 \%, while at position C they decrease
the conductance by 19 \% and 27 \%, respectively. In Ref. \onlinecite{mertig}
it was found that single Fe, Co (and also Ni) defects in a 2x2 infinite Cu
wire decreased the conductance. By analyzing the DOS it was concluded that the
observed reduction of the conductance is due to a depletion of the $s$ states
in the minority band. The above situation is very similar to the case of an Fe
or Co impurity in position C of the point contact considered, even the
calculated drop of the conductance ($\sim$ -20 \% for Fe and $\sim$ -28 \% for
Co) agrees quantitatively well with our present result. Our result, namely,
that Fe and Co impurities at position A increase the conductance can, however,
not be related to the results of Ref. \onlinecite{mertig}. In order to
understand this feature we have to carefully investigate the LDOS calculated
for the point contact.

\begin{figure}[h]
\begin{center}
\includegraphics[width=15cm,clip]{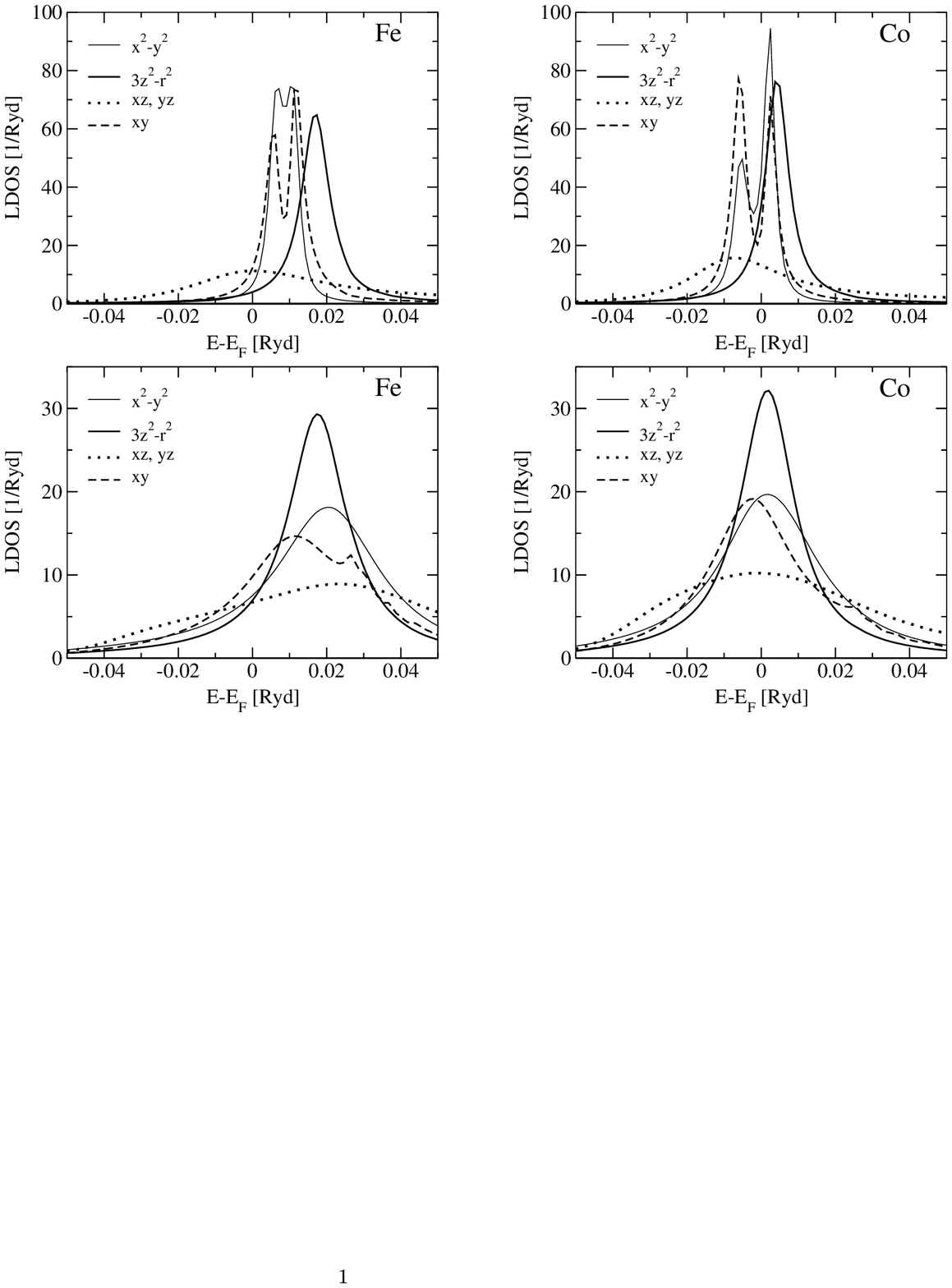}\\[0pt]
\end{center}
\caption{Minority-spin orbital-resolved $d$-like local density of states of Fe
and Co impurities in position A (upper panels) and in position C (lower
panels) of a Au point contact, see Fig. \ref{fig:imp}.}%
\label{fig:dos}%
\end{figure}

In Fig. \ref{fig:dos} we plotted the minority $d$-like LDOS of the Fe and Co
impurities in positions A and C as resolved according to the canonical
orbitals $d_{x^{2}-y^{2}}$, $d_{xy}$, $d_{xz,}$, $d_{xy}$ and $d_{3z^{2}%
-r^{2}}$. We have to stress that this kind of partial decomposition, usually
referred to as the $(\ell,m,s)$ representation of the LDOS, is not unique
within a relativistic formalism, since due to the spin-orbit interaction the
different spin- and orbital components are mixed. However, due to the large
spin-splitting of Fe and Co the mixing of the majority and minority
spin-states can be neglected. As can be seen from Fig. \ref{fig:dos}, the LDOS
of an impurity in position A is much narrower than in position C. This is an
obvious consequence of the difference in the coordination numbers (8 for
position A and 12 for positions C). Thus an impurity in position A hybridizes
less with the neighboring Au atoms and, as implied by the LDOS, the
corresponding $d$ states are fairly localized. Also to be seen is a spin-orbit
induced splitting of about 8 mRyd ($\sim$ 0.1 eV) in the very narrow
$d_{x^{2}-y^{2}}$-$d_{xy}$ states of the impurities in position A. The
difference of the band filling for the two kind of impurities shows up in a
clear downward shift of the LDOS of Co with respect to that of Fe.

\begin{figure}[h]
\begin{center}
\includegraphics[width=15cm,clip]{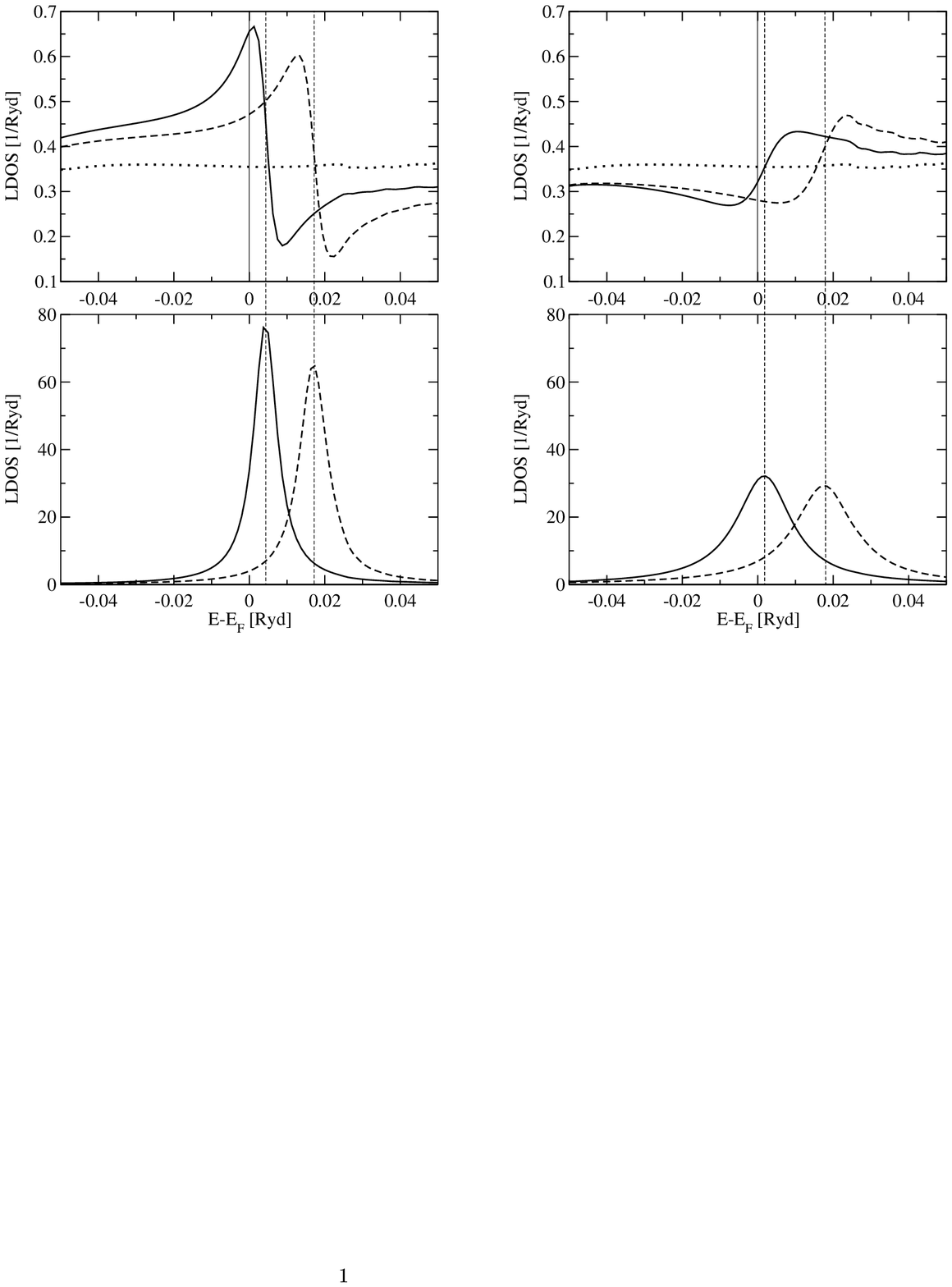}\\[0pt]
\end{center}
\caption{Top left: minority-spin $s$-like local density of states at the
center site of a Au point contact with an impurity at position A, see Fig.
\ref{fig:imp} (solid line: Co, dashed line: Fe). Top right: the same as
before, but with an impuritiy at position C. As a comparison, in both figures
the corresponding LDOS for the pure Au contact is plotted by dotted lines. The
solid vertical lines highlight the position of the Fermi energy. Bottom:
minority-spin $d_{3z^{2}-r^{2}}$ local density of states of the impurities
(solid line: Co, dashed line: Fe) at positions A (left) and C (right).
Vertical dashed lines mark the center positions of the $d_{3z^{2}-r^{2}}
$-LDOS peaks.}%
\label{fig:sdos}%
\end{figure}

In explaining the change of the conductance through the point contact caused
by the impurities in positions A and C, the $s$-like DOS at the center site,
i.e., at the narrowest section of the contact, is plotted in the top half of
Fig. \ref{fig:sdos}. As a comparison the corresponding very flat $s$-like DOS
is shown for a pure Au contact. For contacts with impurities this $s$-like DOS
shows a very interesting shape which can indeed be correlated with the
corresponding $d_{3z^{2}-r^{2}}$-like DOS at the impurity site, see bottom
half of Fig. \ref{fig:sdos}. Clearly, the center positions and the widths of
the $d_{3z^{2}-r^{2}}$-like DOS peaks and those of the respective
(anti-)resonant $s$-like DOS shapes coincide well with each other. This kind
of behavior in the DOS resembles to the case studied by Fano for a continuum
band and a discrete energy level in the presence of configuration interaction
(hybridization) \cite{Fano}. Apparently, by keeping this analogy, in the point
contact the $s$-like states play the role of a continuum and the
$d_{3z^{2}-r^{2}}$-like state of the impurity acts as the discrete energy
level. Since the two kinds of states share the same cylindrical symmetry,
interactions between them can occur due to backscattering effects. It should
be noted that similar resonant line-shapes in the STM I-V characteristics have
been observed for Kondo impurities at surfaces \cite{madhavan,manoharan} and
explained theoretically \cite{ujsaghy}.

Inspecting Fig. \ref{fig:sdos}, the enhanced $s$-like DOS at the Fermi level
at the center of the point contact provides a nice interpretation to the
enhancement of the conductance when the Fe and Co impurity is placed at
position A. As the peak position of the $d_{3z^{2}-r^{2}}$-like states of Fe
is shifted upwards by more than 0.01 Ryd with respect to that of Co, the
corresponding resonance of the $s$-like states is also shifted and the
$s$-like DOS at the Fermi level is decreased. This is also in agreement with
the calculated conductances. In the case of impurities at position C, i.e., in
a position by $a$ = 7.63 a.u. away from the center of the contact, the
resonant line-shape of the $s$-like states is reversed in sign, therefore, we
observe a decreased $s$-like DOS at the Fermi level, explaining in this case
the decreased conductance, see Table \ref{tab:impcond}. Since, however, the
$s$-like DOS for the case of a Co impurity is larger than for an Fe impurity,
this simple picture cannot account correctly for the opposite relationship we
obtained for the corresponding conductances.

\section*{Summary}

By using a Green's function technique based on the embedding scheme of the
multiple scattering theory and the Kubo-Greenwood linear response theory as
formulated by Baranger and Stone \cite{baranger} we studied the conductance of
gold nanocontacts depending on the contact geometry and transition metal
impurities placed at various positions. We performed several numerical tests
that proved the efficiency of our method. In good agreement with experiments
and other calculations we obtained a conductance of 1.1 $G_{0}$ for a finite
linear chain connecting two semi-infinite Au leads. The calculated conductance
for a thicker 2x2 wire, 2.58 $G_{0}$, can be related to a recent result for an
infinite 2x2 chain (3 $G_{0}$) \cite{mertig}. Also in agreement with quantum
mechanical model calculations \cite{Torres} we found a nearly linear
dependence of the conductance on the \textquotedblleft
thickness\textquotedblright\ of the contact. By embedding magnetic transition
metal impurities into a point contact we found both enhancement and reduction
of the conductance depending on the position of the impurities. On analyzing
the local density of states we concluded that the effect of the impurity is
mainly controlled by the interaction of the minority $d$-like and $s$-like
states giving rise to a resonant line-shape (Fano-resonance) in the $s$-like
DOS at the center of contact. We suggest that this line-shape should also be
observed in $I-V$ conductance characteristics providing thus an
\textquotedblleft experimental\textquotedblright\ tool to detect magnetic impurities (even their
position) in a noble metal point contact.

\section*{Acknowledgements}

This paper resulted from a collaboration partially funded by the RTN network
``Computational Magnetoelectronics'' (Contract No. HPRN-CT-2000-00143) and by
the Research and Technological Cooperation Project between Austria and Hungary
(Contract No. A-3/03). Financial support was also provided by the Center for
Computational Materials Science (Contract No. GZ 45.451), the Austrian Science
Foundation (Contract No. W004), and the Hungarian National Scientific Research
Foundation (OTKA T038162 and OTKA T037856).


\begin{table}[h]
\caption{Set-up of various nanocontacts. The table shows the number of Au
atoms in the layers as labelled by C, C $\pm$ 1, etc. in Fig.
\ref{fig:geom-sketch}. Contact 1 refers to Fig. \ref{fig:geom}a. }%
\label{tab:geom}%
{}
\par
\begin{center}%
\begin{tabular}
[c]{|c|c|c|c|c|c|}%
\hhline{------} \hhline{}layer & \multicolumn{5}{c|}{Contact}\\\cline{2-6}%
\hhline{}position & 0 & 1 & 4 & 5 & 9\\
\hhline{------} \hhline{}C$\pm$4 & Full & Full & Full & Full & Full\\
\hhline{}C$\pm$3 & 9 & Full & Full & Full & Full\\
\hhline{}C$\pm$2 & 4 & 9 & 16 & 21 & 25\\
\hhline{}C$\pm$1 & 1 & 4 & 9 & 12 & 16\\
\hhline{}C & 0 & 1 & 4 & 5 & 9\\
\hhline{------} 
\end{tabular}
\end{center}
\end{table}

\begin{table}[h]
\caption{Calculated spin and orbital moments of magnetic impurities placed at
different positions in a Au point contact, see Fig. \ref{fig:imp}. }%
\label{tab:impmom}%
{}
\par
\begin{center}%
\begin{tabular}
[c]{|c|c|c|c|c|}%
\hhline{-----} \hhline{}position & \multicolumn{2}{c|}{$S_{z}[\mu_{B}]$} &
\multicolumn{2}{c|}{$L_{z}[\mu_{B}]$}\\\cline{2-5}%
\hhline{} & Fe & Co & Fe & Co\\
\hhline{-----} \hhline{}A & 3.36 & 2.01 & 0.47 & 0.38\\
\hhline{}B & 3.46 & 2.17 & 0.04 & 0.61\\
\hhline{}C & 3.42 & 2.14 & 0.07 & 0.22\\
\hhline{-----} 
\end{tabular}
\end{center}
\end{table}

\begin{table}[h]
\caption{Calculated conductances of a Au point contact with impurities on
different positions, see Fig. \ref{fig:imp}.}%
\label{tab:impcond}%
{}
\par
\begin{center}%
\begin{tabular}
[c]{|c|c|c|c|}\hline
impurity & \multicolumn{3}{c|}{Conductance $[G_{0}]$}\\\cline{2-4}%
position & Pd & Fe & Co\\\hline
A & \ 2.22 \  & \ 2.67 \  & \ 2.97 \ \\\hline
B & 2.24 & 2.40 & 2.26\\\hline
C & 2.36 & 1.95 & 1.75\\\hline
pure Au & \multicolumn{3}{c|}{2.40}\\\hline
\end{tabular}
\end{center}
\end{table}


\begin{thebibliography}{99}                                                                                               %
\bibitem {mcbj1}C.J.\ Muller, J.M.\ van Ruitenbeek, and L.J.\ de Jongh,
Phys.\ Rev.\ Lett.\ \textbf{69}, 140 (1992)

\bibitem {mcbj2}J.M.\ Krans, I.K.\ Yanson, Th.C.M.\ Govaert, R.\ Hesper, and
J.M.\ van Ruitenbeek, Phys.\ Rev.\ B \textbf{48}, 14721 (1993)

\bibitem {Smit}R.H.M.\ Smit, C.\ Untiedt, A.I.\ Yanson, and J.M.\ van
Ruitenbeek, Phys.\ Rev.\ Lett.\ \textbf{87}, 266102 (2001)

\bibitem {halbritter}A.\ Halbritter, Sz.\ Csonka, O.\ Yu.\ Kolesnychenko,
G.\ Mih\'aly, O.I.\ Shklyarevskii, and H. van Kempen, Phys. Rev. B
\textbf{65}, 045413 (2002); Sz.\ Csonka, A.\ Halbritter, G.\ Mih\'aly,
E.\ Jurdik, O.I.\ Shklyarevskii, S.\ Speller, and H.\ van Kempen, Phys. Rev.
Lett. \textbf{90}, 116803 (2003)

\bibitem {crommie}M.F.\ Crommie, C.P.\ Lutz, and D.M.\ Eigler, Science
\textbf{262}, 219 (1993)

\bibitem {gimzewski}J.K.\ Gimzewski and R.\ M\"oller, Phys.\ Rev.\ B
\textbf{36}, 1284 (1987)

\bibitem {brandbyge}M.\ Brandbyge, J.\ Schi\o tz, M.R.\ S\o rensen,
P.\ Stoltze, K.W.\ Jacobsen, J.K.\ N\o rskov, L.\ Olesen, E.\ Laegsgaard,
I.\ Stensgaard, and F.\ Besenbacher, Phys.\ Rev.\ B \textbf{52} 8499 (1995)

\bibitem {AuPd}A.\ Enomoto, S.\ Kurokawa, and A.\ Sakai, Phys.\ Rev.\ B
\textbf{65}, 125410 (2002)

\bibitem {heemskerk}J.W.T.\ Heemskerk, Y.\ Noat, D.J.\ Bakker, J.M.\ van
Ruitenbeek, B.J.\ Thijsse, and P.\ Klaver, Phys.\ Rev.\ B \textbf{67}, 115416 (2003)

\bibitem {RuitRev}N.\ Agra\"\i t, A.L.\ Yeyati, and J.M.\ van Ruitenbeek,
Phys.\ Rep.\ \textbf{377}, 81 (2003)

\bibitem {brandbyge1}M.\ Brandbyge, N.\ Kobayashi, and M.\ Tsukada,
Phys.\ Rev.\ B \textbf{60}, 17064 (1999)

\bibitem {solanki}A.K.\ Solanki, R.F.\ Sabiryanov, E.Y.\ Tsymbal, and
S.S.\ Jaswal, J.\ Magn.\ Magn.\ Mat., in press

\bibitem {stepanyuk}V.S.\ Stepanyuk, P.\ Bruno, A.L.\ Klavsyuk, A.N.\ Baranov,
W.\ Hergert, A.M.\ Saletsky, and I.\ Mertig, Phys.\ Rev.\ B \textbf{69},
033302 (2004)

\bibitem {mertig}N.\ Papanikolaou, J.\ Opitz, P.\ Zahn, and I.\ Mertig,
Phys.\ Rev.\ B \textbf{66}, 165441 (2002)

\bibitem {opitz}J.\ Opitz, P.\ Zahn, and I.\ Mertig, Phys.\ Rev.\ B
\textbf{66}, 245417 (2002)

\bibitem {landauer}R.\ Landauer, IBM J.\ Res.\ Dev.\ \textbf{1}, 223 (1957)

\bibitem {buttiker}M.\ B\"uttiker, Phys.\ Rev.\ Lett.\ \textbf{57}, 1761 (1986)

\bibitem {kubo}R.\ Kubo, J.\ Phys.\ Soc.\ Jpn.\ \textbf{12}, 570 (1957)

\bibitem {greenwood}D.A.\ Greenwood, Proc.\ Phys.\ Soc.\ London \textbf{71},
585 (1958)

\bibitem {luttinger}J.M.\ Luttinger, in \textit{Mathematical Methods in Solid
State and Superfluid Theory} (Oliver and Boyd, Edinburgh) Chap.\ 4, pp. 157, 1962

\bibitem {butler}W.H.\ Butler, Phys.\ Rev.\ B \textbf{31}, 3260 (1985)

\bibitem {baranger}H.U.\ Baranger and A.D.\ Stone, Phys.\ Rev.\ B \textbf{40},
8169 (1989)

\bibitem {skkr}L.\ Szunyogh, B.\ \'Ujfalussy, P.\ Weinberger, and
J.\ Koll\'ar, Phys.\ Rev.\ B \textbf{49}, 2721 (1994)

\bibitem {rsp-skkr}L.\ Szunyogh, B.\ \'Ujfalussy, and P.\ Weinberger,
Phys.\ Rev.\ B \textbf{51}, 9552 (1995)

\bibitem {bence1}B.\ Lazarovits, L.\ Szunyogh, and P.\ Weinberger,
Phys.\ Rev.\ B \textbf{65}, 104441 (2002)

\bibitem {bence2}B.\ Lazarovits, L.\ Szunyogh, and P.\ Weinberger,
Phys.\ Rev.\ B \textbf{67}, 024415 (2003)

\bibitem {bence3}B.\ Lazarovits, L.\ Szunyogh, and P.\ Weinberger,
Phys.\ Rev.\ B \textbf{68}, 024433 (2003)

\bibitem {palotas}K.\ Palot\'as, B.\ Lazarovits, L.\ Szunyogh, and
P.\ Weinberger, Phys.\ Rev.\ B \textbf{67}, 174404 (2003)

\bibitem {mavropoulos}P.\ Mavropoulos, N.\ Papanikolaou, and P.H.\ Dederichs,
\texttt{cond-mat/0306604} (2003)

\bibitem {Vosko}S.H.\ Vosko, L.\ Wilk, and M.\ Nusair,
Can.\ J.\ Phys.\ \textbf{58}, 1200 (1980)

\bibitem {Torres}J.A.\ Torres, J.I.\ Pascual, and J.J.\ S\'aenz,
Phys.\ Rev.\ B \textbf{49}, 16581 (1994)

\bibitem {Fano}U.\ Fano, Phys.\ Rev.\ \textbf{124}, 1866 (1961)

\bibitem {madhavan}V.\ Madhavan, W.\ Chen, T.\ Jamneala, M.F.\ Crommie, and
N.S.\ Wingreen, Science \textbf{280}, 567 (1998)

\bibitem {manoharan}H.C.\ Manoharan, C.P.\ Lutz, and D.M.\ Eigler, Nature
\textbf{403}, 512 (2000)

\bibitem {ujsaghy}O.\ \'Ujs\'aghy, J.\ Kroha, L.\ Szunyogh, and
A.\ Zawadowski, Phys.\ Rev.\ Lett.\ \textbf{85}, 2557 (2000); O.\ \'Ujs\'aghy,
G. Zar\'and, and A.\ Zawadowski, Solid State Comm.\ \textbf{117}, 167 (2001)
\end{thebibliography}
\end{document}